\documentclass[twocolumn]{optica-article}
\journal{opticajournal}
\usepackage{amsmath,amsfonts}
\usepackage{textcomp}
\usepackage{stfloats}
\usepackage{url}
\usepackage{verbatim}
\usepackage{graphicx}
\usepackage{mathtools}
\usepackage{graphics}
\usepackage{xcolor}
\usepackage{bm}
\graphicspath{{Figures}}
\DeclarePairedDelimiter{\abs}{\lvert}{\rvert}			

\usepackage{siunitx}
\begin{document}
\title{A Simple Method of Evaluating Laser Diode Suitability for Phase-Noise Based QRNG}
\author{Matthias F. Ostner,\authormark{1,*} Innocenzo De Marco,\authormark{1} Christian Roubal,\authormark{1}}
\address{\authormark{1}German Aerospace Center (DLR), Institute of Communications and Navigation, Münchener Str. 20, D-82234 We\ss ling, Germany}
\email{\authormark{*}matthias.ostner@tum.de}

\begin{abstract}
Quantum random number generators (QRNGs) based on semiconductor laser phase noise are an inexpensive and efficient resource for true random numbers. Commercially available technology allows for designing QRNG setups tailored to specific use cases. However, it is important to constantly monitor whether the QRNG is performing according to the desired security standards in terms of independence and uniform distribution of the generated numbers. This is especially important in cryptographic applications. This paper presents a test scheme that helps to assess the acceptable operating conditions of a semiconductor laser for QRNG operation, using commonly accessible methods. This can be used for system monitoring, but crucially also to help the user choose the laser diode which better suits their needs. Two specific quality measurements, ensuring proper operation of the device, are explained and discussed. Setup-specific approaches for setting an acceptance boundary for these measures are presented and exemplary measurement data showing their effectiveness is given. By following the comprehensible procedure described here, a QRNG qualification environment tailored to specific security requirements can be reproduced. 
\end{abstract}

\section{Introduction}
Random numbers are essential for various modern applications, with differing requirements based on context. In quantum key distribution, the aim is to produce independent and identically distributed (IID) numbers, ensuring that the random number outputs are uncorrelated and show uniform probability distribution across the space of potential outputs~\cite{Gentle.2012, Bassham.2010}. Quantum random number generators (QRNGs) utilize random quantum phenomena, such as the detection of single photons after they pass through a beamsplitter, as their entropy source, aligning with the probabilistic principles of quantum theory~\cite{Jennewein.2000, Rarity.1994}.
QRNG technology as a whole has advanced significantly, and its applications are now expanding beyond quantum communication domains.
It's no longer confined to its main application being a subsystem to quantum key distribution devices, but is now being utilized in various sectors of the general market, with QRNG chips being installed in smartphones and self-driving vehicles to improve their security against external attackers.

\begin{figure*}
    \centering
    \includegraphics[width=\linewidth]{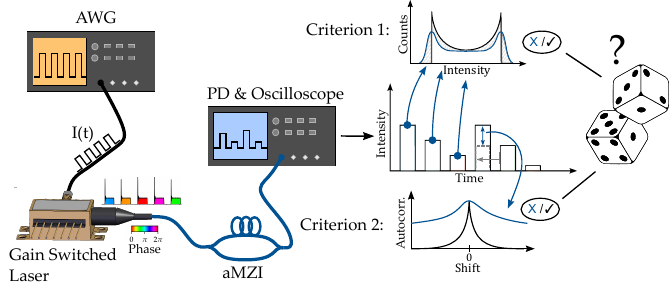}
    \caption{Experimental QRNG setup and methods used in this paper. A semiconductor laser is operated in Gain Switching mode with the help of an arbitrary waveform generator (AWG). The driving current $I(t)$ and chip temperature could be varied. The generated laser pulses show phase randomization indicated by the various colors under the pulse envelopes. The pulse train interferes with a delayed version of itself in an asymmetric Mach-Zehnder-Interferometer (aMZI), resulting in pulses of randomized intensity. These are measured with a photodiode (PD) and an oscilloscope. The intensities extracted from these pulses are scrutinized in terms of their statistical properties with Criterion 1, the normalized statistical distance giving the difference of the measured intensity distribution (blue) with respect to an ideal arcsine (black). Criterion 2 is the autocorrelation of the raw pulse train and indicates how similar the pulse train is to a shifted version of itself, shown with gray dashed lines. Ideally, the autocorrelation drops already for low shifts (black curve). If both criteria are fulfilled, the built QRNG setup is assumed to produce true random numbers.}
    \label{fig:overview_drawing}
\end{figure*}

A fast and effective type of QRNG is based on phase fluctuations in gain-switched semiconductor lasers, originally presented by Jofre et al. in~\cite{Jofre.2011}. The general idea behind this approach is that when the photon density in the laser medium is low, e.g. when the laser is switched off, spontaneously emitted photons randomize the phase of the cavity field and this phase randomization leads to random interference intensities of subsequent pulses emitted in the lasing phases of the gain-switched laser. This interference is achieved with an asymmetric Mach-Zehnder-Interferometer (aMZI).

QRNGs based on this concept have been extensively studied and developed.
Since the first demonstration in 2011~\cite{Jofre.2011}, major improvements have been made.
In particular, much work has been done to improve the key rates~\cite{Yang.2023}, miniaturize the optical systems~\cite{Abellan.2016, Roger.2019, Marangon.2024}, and develop new hardware and software-based post-processing techniques to enable real-time processing and generation~\cite{Guo.2024, Li.2024}.

The model developed by Henry et al.~\cite{Henry.1986} describes the cavity field phase diffusion due to spontaneous emission leading to a Gaussian-like phase distribution between subsequent pulses. Since for interference intensities only relative phase values between $[-\pi,\pi]$ are relevant, the Gaussian phase distribution gets projected onto this interval and produces a uniform phase distribution across it, after typical off-times in the order of \SI{e-10}{\second}~\cite{Septriani.2020}. This necessary off-time, strongly dependent on the bias current applied to the laser diode, limits the pulse repetition frequency to a few \SI{}{\giga\hertz} and therefore also the random number generation rate.

The interference intensities caused by a uniform underlying phase distribution are not uniformly distributed but describe an arcsine distribution due to the cosine-dependence of the interference intensity.
Due to this distribution, the resulting raw pulse interference intensity values obtained from digitization with an ADC as described in Section \ref{sec:setup} are also not uniform. In order to get a close-to-uniform distribution of the QRNG output numbers and clear the output string of numbers from classical and non-random noise contributions, one  needs to perform a randomness extraction step. This is necessary even when digitizing the pulse intensities with comparators, yielding close-to-uniform distributions directly, due to the classical noise \cite{Shakhovoy:20}. There are several extraction methods suitable for cryptographic applications, one example is the seeded Toeplitz extractor~\cite{Dodis.2020} which reduces the bit length of each output but thereby increases the uniformity of the distribution and decreases the amount of cryptographically insecure numbers to a point sufficient for the specific requirements of a given application. This last step is not treated in this paper, so other sources can be considered for more detail, e.g.~\cite{Abellan.2014}.

The design and experimental realization of such a QRNG, as outlined in Section~\ref{sec:setup}, is comparatively straightforward. Utilizing a custom-made setup allows customization to specific security requirements, enhancing trustworthiness and minimizing dependence on external manufacturers, while also potentially optimizing system design by conserving resources such as weight, space, or power.

However, there are complications to be aware of. Although the entropy source of such devices is based on unpredictable quantum processes, which makes them ideal for true random number generation, potentially predictable classical noise influences the outputs of these generators, especially in the measurement and digitization steps of the procedure, but also in the optical output due to effects like jitter and chirping~\cite{Shakhovoy.2021}. Randomization of the output can also be hindered by backreflection of light into the laser cavity~\cite{Shakhovoy.2025} or by attacks on these devices e.g. by injecting electrical signals into the measurement stage through RF antennas~\cite{Smith.2021}. Therefore, a requirement for secure generation of random numbers is that the used device can be trusted and that the user is aware of the various sources of noise influencing the QRNG output. This requires a testing framework for the QRNG to make sure that the QRNG is performing according to the defined requirements. Examples of approaches to increase the security of a QRNG via the Min-Entropy formulation exist and can be found e.g. in~\cite{Rude.2018} and~\cite{Mitchell.2015}. However, gaining the necessary detailed knowledge about classical contributions and incorporating it into the randomness extraction procedure can be experimentally and mathematically challenging and time-consuming.

An easier approach towards quality assurance of pulsed, phase-noise-based QRNGs is presented in this paper. The goal of this work is not to quantify how much information can be retrieved from raw intensity values but to propose a simpler qualification procedure that allows for discriminating between operating conditions that allow for QRNG operation and those that don't. 
This proves useful in monitoring the performance of a running QRNG, similar to previously reported work~\cite{Marangon.2024}, but crucially it also provides a full qualification framework to assess whether or not a laser is suitable to be used as a QRNG in the first place. While in particular the relationship between the laser diode driving current and the capability of such a QRNG to produce quantum random numbers was studied in detail before \cite{Shakhovoy.2023}, the generality of the presented qualification framework is a valuable addition to existing approaches.
The main idea behind this work is to make it easier for experimentalists to select an appropriate laser for their application.
Performance of a laser diode can be affected by many factors, one example being electrical bonding and packaging limiting the modulation bandwidth.
This work provides a mean to verify whether these limiting factors allow the device to be used as intended.

In the following, a comprehensive and transparent framework consisting of two quantitative quality measures is presented.
Our framework allows for easy implementation in various circumstances. An exemplary set of boundary conditions is given to ensure independence of the generated random numbers and the uniformity of the underlying phase distribution, the preconditions for IID number generation. Furthermore, methodological requirements and risks are explored. Finally, experimental data is presented which proves the applicability of the defined test procedure.

\begin{figure}
    \centering
    \includegraphics[width=1\linewidth]{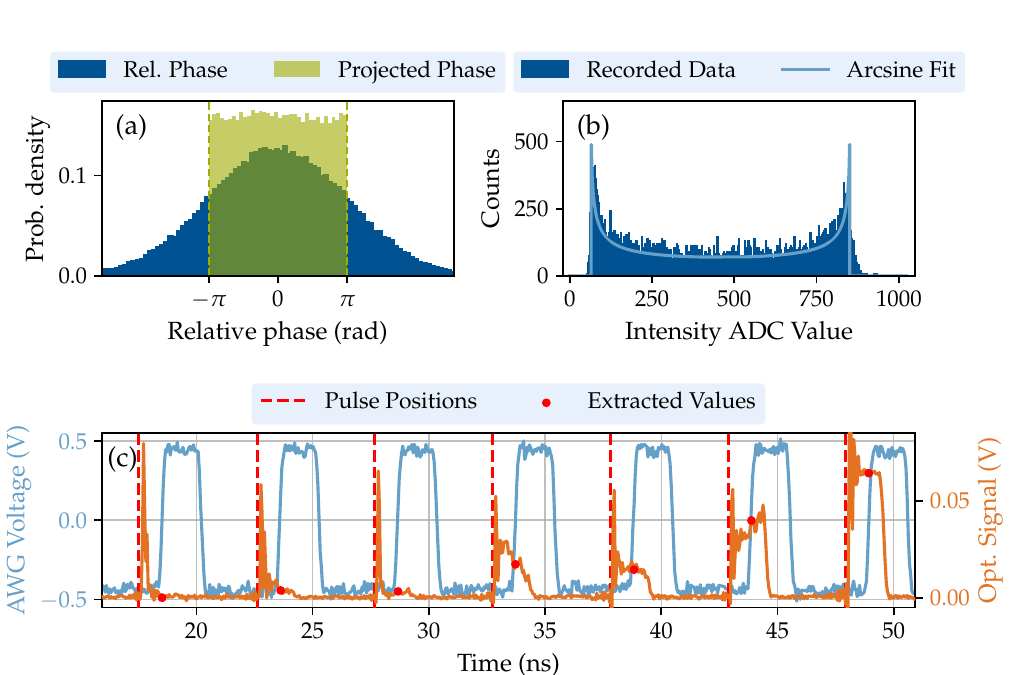}
    \caption{(a) A Gaussian distribution of the relative phase between two subsequent laser pulses evolves through phase diffusion. Projected to the phase space relevant for interference, it gives a uniform distribution after some time. (b) A recorded pulse intensity histogram and a fitted ideal arcsine distribution resulting from perfect phase randomization. (c) Electrical and optical pulse trace detail recorded with the described setup. The automatically extracted intensity values are marked with red dots and contribute to the intensity distribution above.}
    \label{fig:phase_drift}
\end{figure}

\section{Experimental QRNG Setup}
\label{sec:setup}
The setup for the custom QRNG evaluated in this manuscript is similar to the one originally proposed by Jofre et al.~\cite{Jofre.2011} and consists of a gain-switched semiconductor laser module connected to a fiber-based aMZI. All fiber components used for the setup are made of polarization-maintaining optical fiber. The current pulses driving the laser consist of a constant bias current from the laser driver and a superimposed non-amplified square signal from a $\SI{6}{\giga\siemens\per\second}$ arbitrary waveform generator (AWG) applied via a Bias-T. The produced light pulse train interferes with a copy of itself delayed by $\Delta\tau$ and the interference intensity is measured with a $\SI{50}{\giga\hertz}$ photodetector and an oscilloscope with an ADC bandwidth of $\SI{8}{\giga\hertz}$, a sampling rate of \SI{20}{\giga\siemens\per\second}, and a vertical resolutionn of $10$ bits. The measured intensities are recorded as a raw string of random bits. 
The acquisition of one intensity value per pulse is synchronized via the electrical signal of the AWG combined with a measured delay between the electrical and optical pulses. The measurement timing was set to obtain data points after the relaxation oscillation regime, if the pulse length permitted.
A sketch of the setup is presented in Figure \ref{fig:overview_drawing} where also the procedure to evaluate the measured pulse intensities, explained in Section \ref{sec:Methods}, is visualized.
Three laser diodes from different manufacturers were scrutinized as QRNG sources and $5000$ different combinations of operating parameters per laser were evaluated for their ability to generate quantum random numbers according to the two criteria explained in sections~\ref{subsec:stat_dist} and~\ref{subsec:autocorrelation}. 
Relevant specifications of the three lasers are shown in Table \ref{tab:laser_specs}.
\begin{table}
\centering
    
    \caption{Relevant laser specifications for the three scrutinized models taken from their respective datasheet.}
    \label{tab:laser_specs}
    \begin{tabular}{lccc}
    \hline
    Typical Parameter Value                                                                       & Laser 1 & Laser 2 & Laser 3       \\ \hline
    Center Wavelength {[\SI{}{\nano\meter}]}   & $1550$  & $1540$  & $1550$        \\
    Laser Linewidth {[\SI{}{\nano\meter}]]}     & $0.1$   & $-$     & $\SI{8e-6}{}$ \\
    Threshold Current {[\SI{}{\milli\ampere}]} & $6$     & $9$     & $8-20$        \\
    Operating Current {[\SI{}{\milli\ampere}]} & $20$    & $50$    & $75$          \\
    Output Power {[\SI{}{\milli\watt}]}        & $2$     & $6$     & $10$          \\
    Chip Temperature {[\SI{}{\celsius}]}                      & $15-35$ & $15-45$ & $15-35$\\
    Modulation Bandwidth {[\SI{}{\giga\hertz}]} & $<1$ (Bias-T) & $>15$ & $>10$ \\ \hline
\end{tabular}
\end{table}

All of them are DFB lasers with a center wavelength around $\SI{1550}{\nano\meter}$. 
For each parameter set, $10000$ pulses were recorded and the intensity values were evaluated with both criteria for their suitability for QRNG application. 
The varied driving parameters were the chip temperature $T$, the duty cycle $\text{DC}$ of the pulses, the pulse peak current $I_m$, and the composition of the pulses determined by a modulation depth parameter $\text{MD}$ determining how high the constant bias current and the modulation signal amplitude were. When $\text{MD}$ surpasses $0.5$, the modulation signal amplitude is greater than the constant bias so that reverse biasing is achieved in the off-phases of the lasers.

The used aMZI was built by splicing two fiber-based 
$50/50$ couplers together, with one arm being \SI{1}{\meter} longer than the other, resulting in a delay of $\Delta\tau=\SI{5.0678 \pm 0.0584}{\nano\second}$. This delay corresponds to an ideal pulse repetition frequency of $\SI{197.32}{\mega\hertz}$ leading to perfect overlap of the pulses. An average splitting ratio of $\SI{51.50}{}$/$\SI{48.50}{}$ $\pm 2.56$ was measured.
The aMZI is a very important component in this system and it was designed to match the application requirements. In this case the AWG sampling rate defined the necessary delay.

The system phase noise was measured under various laser operating conditions with the three different, commercially available laser types. 
The reason for using different lasers is to compare their behavior and quality in order to show how the acceptable operating window changes depending on the device. This is presented in more detail in Appendix~\ref{app:comparison}. Driving the lasers in CW mode above threshold, the intensity fluctuation of the system was measured by taking intensity measurements every $\SI{5}{\nano\second}$, reflecting the distance between subsequent QRNG pulses, and converted to phase noise values by using the well-known relation between interference intensity and phase difference $I(\Delta\Phi)$ and the maximum and minimum achievable intensities in this operating mode. The worst-case gave a Gaussian distribution of the phase fluctuations between two subsequent pulses with a width of $\sigma=\SI{36.33\pm0.41}{\degree}$ which constitutes substantial phase noise. However, it is far from being strong enough to give a uniform distribution across the whole interval $[-\pi,\pi]$, which is sufficient to distinguish between driving conditions that enable phase randomization caused by spontaneous emission and such that don't. Therefore, the setup is suitable to investigate the QRNG quality measures introduced in Chapter~\ref{sec:Methods}.

\section{Quantitative Decision Rules for QRNG Operation}
\label{sec:Methods}
In order to qualify certain laser operating conditions as phase randomizing and therefore suitable for quantum random number generation, quantitative analyses of the system and the measurement results are needed. There are concepts how to approach this task. One is a Min-Entropy formulation to estimate how much quantum randomness can be extracted from the numbers obtained. The most basic formulation of the Min-Entropy is
\begin{equation}
    \label{eq:min_entropy}
    H_{min}(X) = -\log_2{(\max_{1\leq i \leq k} p_i)}\text{,}
\end{equation}
where $X=\{x_1,x_2,...,x_k\}$ is a discrete random variable and $p_i$ is the probability of outcome $x_i$. This measure gives the target bit length at the end of the final extraction process. Depending on the imbalance in the initial intensity distribution, more or less information can be obtained where the maximum is achievable with a perfectly uniform distribution $p_i=1/k$. This formulation is used e.g. in~\cite{Lovic.2021} as a QRNG quality measure but it does not take into account that classical noise contributes to the measurement and broadens the underlying phase distribution so that the phase drift is not only caused by quantum processes. This lowers the security of the obtained random numbers. 
Also, when the classical noise smears out the arcsine distribution, as visualized in Figure~\ref{fig:phase_drift}, the Min-Entropy increases although one can tell less about the underlying phase distribution. 
One approach to solve this issue is to adapt the Min-Entropy formulation towards a conditional Min-Entropy requiring detailed knowledge about the system components. This is done in detail in~\cite{Abellan.2014, Mitchell.2015} and applied for example in~\cite{Rude.2018}. 
Getting this knowledge about the individual noise contributions is a laborious task, especially when environmental circumstances vary. In order to overcome the weaknesses of the Min-Entropy formulation and find a more practical way to the qualification of QRNGs based on the concept described, an approach via the statistical distance is described in Section~\ref{subsec:stat_dist} to qualify sufficient close-to-uniform phase randomization or identical distribution. Such an approach relies on the fact that classical noise will broaden the arcsine distribution; there still is the need to know the overall noise in the detection system, but measuring this is less complex compared to an estimation for each individual source of noise.
In order to test a QRNG output for independence of the raw outcomes, additionally, an autocorrelation criterion is presented. These two criteria together, if applied correctly, constitute a simple testing environment for QRNGs generating IID numbers.

\subsection{Statistical Distance}
\label{subsec:stat_dist}
The first criterion mentioned is the statistical distance between the measured intensity histogram $I~=~\{n_0,n_1,\dots,n_{1023}\}$ and a fitted arcsine distribution $A~=~\{a_1,a_2,\dots,a_{1023}\}$. The $n_i$ indicate the number of intensity measurements at the ADC value $i$ between $0$ and $1023$ and the $a_i$ represent the expected measurements in the respective ADC bin in the case of an ideal arcsine distribution. The quantity is assumed to be suitable as it quantifies the underlying phase uniformity by inferring it from the closeness between the measured intensity histogram and the ideal one that would follow from a uniform underlying phase. 
The value is calculated as follows: 
\begin{equation}
\label{eq:norm_stat_dist}
d_{stat} = \frac{1}{2 N_I}\sum_{i=0}^{1023} \abs{a_i-n_i}
\end{equation}
The center value of each arcsine bin should be taken so that there is no divergence at the boundaries of the definition range.
For each measurement, $N_I=\sum_{i=0}^{1023}n_i$ is the overall amount of intensities measured in one run. 
It holds that $N_I = \sum_{i=0}^{1023}a_i$, so for the fitting of the arcsine distribution the area under the arcsine corresponds to the total amount of measurements taken. 
In the actual experiment, the statistical distance is measured by recording an intensity histogram and fitting an ideal arcsine with the same area as the actual distribution to it while minimizing the statistical distance.

In~\cite{Marangon.2024}, the statistical distance, or total variation distance as they call it, was already used to monitor their QRNG performance. The difference in the approach is that they evaluated the statistical distance between a current distribution and an initially recorded reference distribution, which makes sense in a scenario where the QRNG has been already characterised and the best possible curve has been measured. In our approach, the ideal arcsine is fitted to minimize the distance to the measured intensity distribution. This makes the approach more robust against changes in the ADC dynamic range fraction used by the recorded intensities, which might be caused e.g. by a varying laser output power. Additionally, there is no previously recorded histogram, which needs to be obtained in the first place, necessary to qualify the measured histograms but a numerical approach can be taken to define the boundary for the criterion.

Before describing this numerical approach to find a suitable boundary condition, a theoretical value for the statistical distance under ideal conditions and for one extractable bit from the 10-bit ADC is established to verify the plausibility of the the numerically obtained value. Under the assumption of a perfect device without noise and with $N_I = 10000$ measurements spread over $\SI{50}{\percent}$ of the ADC, representing the worst-case conditions for the system used here, this value is calculated in the following way. First, the Min-Entropy defined in Equation \eqref{eq:min_entropy} is used to find that for one extractable bit, the maximum probability for a single bin must be $p_{max}\leq 1/2$. If one starts with an arcsine distribution, takes counts from other bins, and rearranges them into a specific bin $x$ until its height reaches half of the total amount of measurements, the statistical distance consequently increases. It increases until it reaches a statistical distance of $\leq 0.5$, because about half of the measurements are misplaced with respect to the ideal distribution. Suppose that the bin with index $x$ contains half of the total measurement points after rearrangement, so that $n_x~=~N_I/2$. Then the difference in counts $n_x - a_x$ is taken from the other bins so that the statistical distance gives:
\begin{eqnarray}
        d_{1} &&= \frac{1}{2N_I}\left(\abs{(n_x-a_x)}+\sum_{i=0, i\neq x}^{1023} \abs{a_i-n_i}\right)\nonumber\\&&=\frac{1}{2N_I}\left(\left|\frac{N_I}{2}-a_x\right|+\left|\frac{N_I}{2}-a_x\right|\right)\\&&= \left|\frac{1}{2}-\frac{a_x}{N_I}\right|\nonumber
\end{eqnarray}
To get the lowest value for $N_I~=~10000$ with reasonable parameters from actual measurements with the used setup, $a_x$ should be maximal. This is reached by using the first or last - so the highest - bin of the narrowest reasonable arcsine, spanning half of the 10-bit ADC dynamic range. This value for $a_x$ is calculated with the probability density function of the arcsine distribution to give $a_x~=~281$. This results in a theoretical value of $d_1~=~0.472$ for this ideal case which must be undercut to have at least one bit to extract. From the real setup more information should be extracted. Therefore, the statistical distance boundary should be lower than this.
To determine a more restrictive boundary tailored for the used setup, including its noise information and assuming non-perfect phase distribution, a numerical approach was used which indeed gave a lower boundary value.

For that numerical approach, the phase diffusion width $\sigma_\Phi$ necessary for a certain application has to be determined first. Then, with that width, the intensity output of interference measurements can be simulated. The system's overall noise needs to be measured and convolved with the generated histogram. Finally, fitting the arcsine with minimal statistical distance gives the boundary value for the QRNG below which one can accept the operating conditions. The main advantage of this approach is that it is easier to implement than the detailed Min-Entropy adaptions mentioned above but nevertheless takes into account all experimental realities. 

Determining the minimum width of the phase drift distribution between two pulses is more or less arbitrary and depends on the security of the random numbers that is desired. A study of the connection between security in QKD protocols and phase drift width is found in~\cite{Kobayashi.2014} together with numerical examples. It is important to know that the amount of quantum randomness extractable from the raw data depends on the phase drift width. A detailed discussion on randomness extraction for the  cases $\sigma_\Phi > 2\pi$ and $\pi \leq \sigma_\Phi \leq 2\pi$ is presented in ~\cite{Shakhovoy.2023}. Due to this dependence, a trade-off between security and pulse generation frequency, so QRNG speed, has to be made as a wider phase drift is achieved by longer off-times of the laser. More precisely, the phase drift width is linearly proportional ($\sigma_\Phi^2\propto t$) to the elapsed time according to theory~\cite{Henry.1986}. A reasonable threshold for the Gaussian phase distribution width was estimated in~\cite{Septriani.2020} to be $\sigma_\Phi^2=(0.8\pi)^2$ and taken as the goal in this study. Given the relative phase noise $\sigma_\Phi$ of the used setup described above, the target phase drift is increased to $\sigma_\Phi^2=(0.825\pi)^2$, in order to guarantee that the actual quantum phase drift is exceeding the threshold.

In a next step, an ideal arcsine was fitted to the generated histogram, and the statistical distance was calculated. This was done for different values of the intensity noise and for different fractions of the recorded intensity values within the dynamic range of the ADC. In the experiment, different dynamic range settings had to be used, as the varied driving current led to varying intensity output. In the selected range of settings, the intensity noise of the detection system was measured and found to be in the range of \SIrange{0.43}{3.0}{\percent} of the maximum intensity at that setting. As the oscilloscope range cannot be set in arbitrarily fine steps, the measured intensities usually do not cover the whole ADC range, instead they typically take up \SIrange{50}{85}{\percent}.

\begin{figure}
    \centering
    \includegraphics[width=1\linewidth]{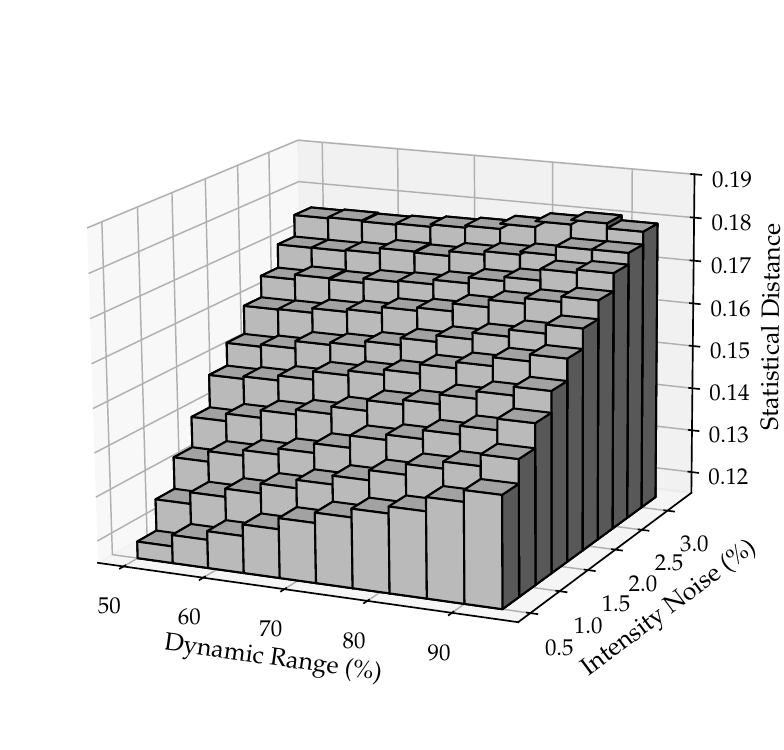}
    \caption{Simulated statistical distance for varying ADC dynamic range fractions taken by the recorded intensity values and varying intensity noise in the system in percent of the ADC range. There is monotonous dependence of the statistical distance with respect to both quantities. The mean $d_{stat}~=~0.155$ serves as the boundary.}
    \label{fig:stat_dist_boundary}
\end{figure}

These two parameters, the noise and the ADC fraction, were scanned with the above procedure and a mean statistical distance of $d_{mean}=0.155$ was found for the typical operating conditions of the used device. This value is used as the boundary for the statistical distance criterion, so that QRNG operation is assumed for measured values $d_{stat}\leq d_{mean}$. From the simulated data shown in Figure~\ref{fig:stat_dist_boundary}, one can see that for increasing noise and intensity fraction, the statistical distance obtained seems to increase almost linearly. The increase with noise is desirable, as this, contrary to the increase of the Min-Entropy, leads to a rejection of generated numbers if the intensity distribution is so different from an arcsine that one cannot infer a sufficient underlying phase randomization. The decrease for lower dynamic range values bears a risk that one needs to be aware of. Due to the discretization of the measurement values and therefore the intensity distribution, less dynamic range means less resolution of the two compared distributions. In the extreme case, where each measurement falls within only one ADC bin, the statistical distance would be zero, as also for a very slim arcsine distribution all values fall within this bin. This case would also cause the autocorrelation described below to be zero for any shift, so the whole testing procedure would break down. To prevent this, the lower boundary for the width of the fitted arcsine needs to be limited to the lowest expected dynamic range of the used device combination.

\begin{figure}
    \centering
    \includegraphics[width=1\linewidth]{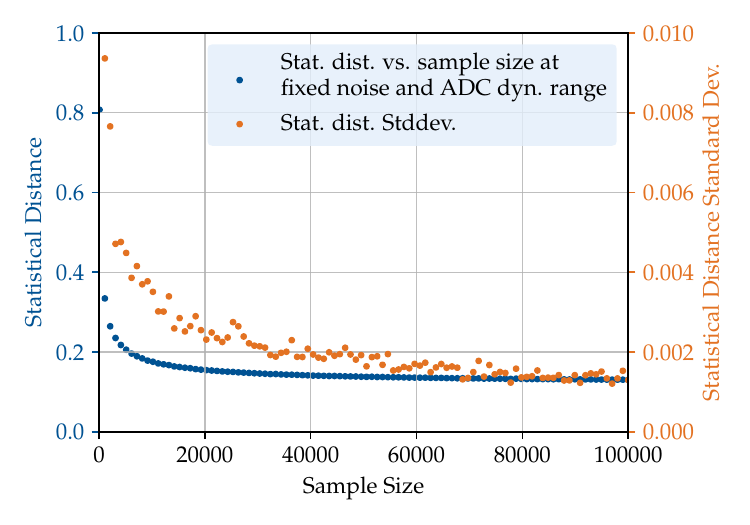}
    \caption{Simulated Statistical distance and its standard deviation at a fixed noise value of \SI{1.5}{\percent} and an assumed ADC dynamic range fraction of \SI{50}{\percent}. With increasing sample size, both values decrease and converge. This emphasizes that simulations have to match the experimental conditions.}
    \label{fig:stat_dist_vs_sample_size}
\end{figure}

Regarding this numerical approach it is also important that it reflects the sampling strategy of the actual verification measurements in terms of sample size. In Figure~\ref{fig:stat_dist_vs_sample_size} one can see that for identical values of noise and ADC dynamic range the statistical distance varies for varying sample sizes and eventually converges. One should be aware of that behavior and perform the boundary estimation with the same parameters as the actual qualification measurements. 
One should also be aware that the standard deviation of the statistical distance measurements depends on this sample size. In the same figure the exponential decrease in fluctuations with increased sample size is shown. This is something to be aware of when doing the trade-off between sampling speed and accuracy of the measurements. It should be noted, that this is not a downside of the presented measure but fundamental statistical fluctuation caused by the measurement.

\subsection{Autocorrelation}
\label{subsec:autocorrelation}
Besides the criterion for the statistical distance, another important quantity was chosen to qualify the randomness of the raw extracted pulse intensities and therefore the generated random numbers. This quantity is the autocorrelation $\Gamma_d$ of the sequence of $N_I$ extracted numbers $X_i$, 
\begin{equation}
\label{eq:autocorrelation_coeff}
    \Gamma_d = 
    \frac{1}{N_I}\sum_{i=1}^{N_I} X_i\cdot X_{i+d} -\overline{X}^2\text{.}
\end{equation} 
The autocorrelation is a measure of how similar the sequence of random numbers is to a shifted version of itself. If the autocorrelation for a certain shift is high, information about later numbers can be gained by knowing previous ones which makes them predictable. High autocorrelation values can therefore disqualify the independence of generated random numbers. Vice versa, a low enough autocorrelation is an indicator for independence of the QRNG output.
To compare the autocorrelations for different signals, one can calculate a normalized autocorrelation coefficient~\cite{Gajic.2003} $C_d = \frac{\Gamma_d}{\Gamma_0}$. This coefficient will take a value between $[-1,1]$. In this formalism, $C_0=1$ is always true and this value is the reference for the absolute values of the autocorrelation coefficients given in \SI{}{\decibel} from here on. 

For a random signal, a rapid drop of the autocorrelation coefficient is expected and necessary already for low shifts. For the experiment in which the laser parameter space is explored, the autocorrelation coefficient value for $d=1$ is taken as the second criterion to qualify certain operating conditions as phase randomized. As in the case of the statistical distance, a boundary value can also be chosen for the autocorrelation, above which a string of random numbers is rejected. This boundary should be chosen adequately for a given need of security in a specific application. For this analysis, a boundary was defined by measuring the autocorrelation coefficient for $d=1$ for intensity values obtained from CW operation of the lasers above threshold with the interferometer connected.

The measured coefficients for different settings and laser models ranged from  \SI{-15.42 \pm 0.10}{\decibel} to \SI{-0.70\pm0.01}{\decibel}. This means that if the extracted random numbers in gain-switching mode show a lower $C_1$ than \SI{-15.52}{\decibel}, they are more independent than the numbers extracted from CW light with superimposed noise. The boundary on the autocorrelation coefficient $C_1$ for the demonstration measurements was chosen to be $C_1\leq\SI{-18.52}{\decibel}$, so half the CW value. This boundary was used as the conducted measurements should demonstrate the identification of QRNG operation. Achieving a higher independence of the outcomes than that produced by non gain-switching operation is sufficient in this case.

\begin{figure}
    \centering
    \includegraphics[width=1\linewidth]{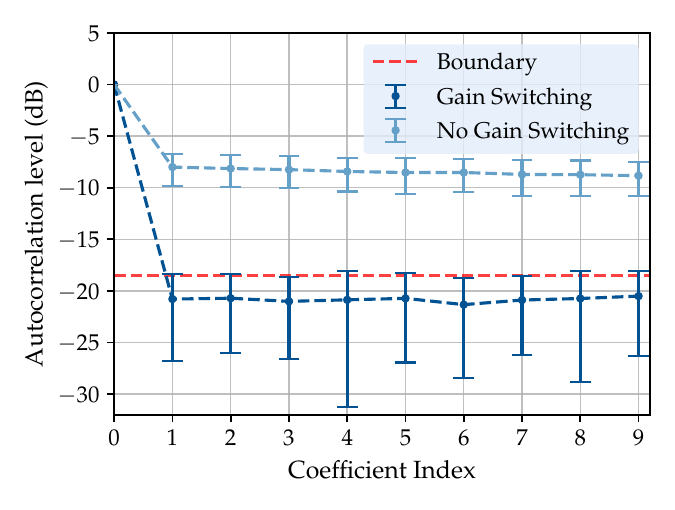}
    \caption{Measured autocorrelation coefficients $C_i$ and deviations under QRNG and non-gain-switching conditions. Both datasets were recorded with Laser 3. In QRNG operation, the coefficients are below the boundary and do not decrease with increasing index $i$. Under non-gain-switching conditions, the means decrease at higher indices.}
    \label{fig:laser3_autocor}
\end{figure}

It should be noted that this choice only suits this application. In a real application, the boundary condition should be chosen so that the security needs are matched. Furthermore, taking into account the higher autocorrelation coefficients $C_i$ with $i>1$ is advisable in order to improve the protection against attacks like the one described in~\cite{Smith.2021}, where RF signals were injected into the measurement setup, or device failures. In the lab setting none of these concerns were relevant and measuring the first $10$ autocorrelation coefficients for a certain laser setting producing gain-switching conditions showed that in the case of randomization, the coefficient means do not decrease further when increasing the index $i$. For non-randomizing conditions they do. This agrees with the assumptions about the autocorrelation behavior with respect to different levels of coherence between the pulses. A comparison between different operating conditions is presented in Figure~\ref{fig:laser3_autocor}.

When comparing the autocorrelation measurements with rate equation model simulations of the experiment, an initial discrepancy appeared. When the model did not contain a slow phase drift present in the actual setup, the autocorrelation values were underestimated in the simulation model under non-gain-switching conditions. This lead to the acceptance of parameters which were excluded in the actual experiment because they did not enable gain-switching and therefore sufficient phase randomization. 
Because of the used form of the autocorrelation coefficients, shown in equation \eqref{eq:autocorrelation_coeff}, even very stable pulse intensities showing a high autocorrelation can give low coefficients if the variation of the pulse intensities around their mean is very small. Then, the autocorrelation coefficients are sensitive to small variations in pulse intensity leading to results not compatible with the defined boundary.
In order to avoid a misbehavior in actual use of the criterion, one has to make sure that the boundary value is actually determined with the used QRNG setup so that all factors influencing its behavior are taken into account.
A second way would be to adapt the definition of the autocorrelation, e.g. by using the ADC range median as a static mean or by normalizing the autocorrelation with the range of obtained intensity values. Both adaptations were not tested, but they are mentioned as ideas if the definition given here should not be suitable to the reader's device properties.

\section{Validation Measurements}
\label{sec:measurements}
The suggested criteria were tested by validating the three lasers mentioned in~\ref{sec:setup} for QRNG operation through a series of measurements.
The lasers' operating parameter ranges explored both gain-switching and non-gain-switching operation, in order to fully analyse the boundary conditions between random and non-random generation.
This set of measurements works as an exclusion and selection criterion: when developing a QRNG, it is important to select the right components. 
Lasers which do not show a wide enough operating window, based on these measurements, cannot be trusted for effectively generating quantum random numbers, and so they should not be used.
The width of the boundary conditions has to be adjusted depending on individual requirements.
This is the first step of the testing stack: if, for example, the application does not require very high generation rates, the statistical distance can be adjusted for a higher threshold, which would yield lower data rates as the resulting min-entropy would be lower, but would in turn expand the pool of laser diodes that can be usable, opening the door to using cheaper devices.
On the other hand, if the highest possible data throughput is necessary, a laser needs to satisfy much stricter criteria to ensure that the laser is indeed capable of operating at high performance.

\begin{figure}
    \centering
    \includegraphics[width=.75\linewidth]{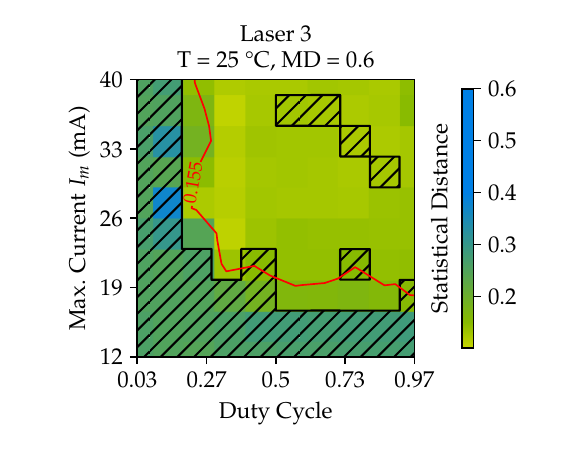}
    \caption{Visual evaluation of the statistical distance and autocorrelation criteria for a given operating parameter plane for Laser 3. Temperature and modulation depth are fixed, the pulse current and duty cycle are varied. The statistical distance criterion is fulfilled below the boundary of $0.155$ indicated by the red line, the autocorrelation is sufficient for parameters which are not shaded. The excluded points in the top right are caused by autocorrelation fluctuations. Such measurement data must be discarded.}
    \label{fig:laser3_plot_dc_vs_Im}
\end{figure}

An exemplary plane in the four-dimensional parameter space explored for Laser 3 is shown in Figure~\ref{fig:laser3_plot_dc_vs_Im}.
The plot shows that both the statistical distance and the autocorrelation criteria with their respective boundaries distinguish between parameters that allow for QRNG operation and those that don't, which is their main purpose.
From analyzing this plane, one can make multiple observations confirming that they reflect expected QRNG behavior and monitor the QRNG performance.
Most importantly, there are operating conditions expected to be rejected, like non-gain-switching ones, and they are rejected by both criteria. For this laser, the exclusion areas of both criteria resemble each other except for some outliers in the upper right region which are caused by fluctuations of the autocorrelation values depicted in Figure~\ref{fig:laser3_autocor}. 

The mentioned fluctuations get more and more relevant for the rejection of measurements the lower the boundary is chosen but they are not problematic. All they mean is that in order to produce good random numbers, some measurement sets should be discarded if their autocorrelation is too high.
Too low of a driving current leads to the rejection of measured values. The optical output during the on-phases must be high enough to produce a good signal-to-noise ratio (SNR), otherwise the arcsine intensity distribution gets buried under the system noise and no phase randomization can be inferred from the smeared-out histogram. This effect is most prominent in Laser 2 which has lower optical output power compared to the other models. In Figure~\ref{fig:dc_vs_max_current_comparison}, one can see the effect, namely that data generated by Laser 2 is only accepted when driving the laser at very high currents.
This also means that just surpassing the threshold current is not sufficient, a certain overshoot depending on the system is necessary to pass the test.

\begin{figure}
    \centering
    \includegraphics[width=1\linewidth]{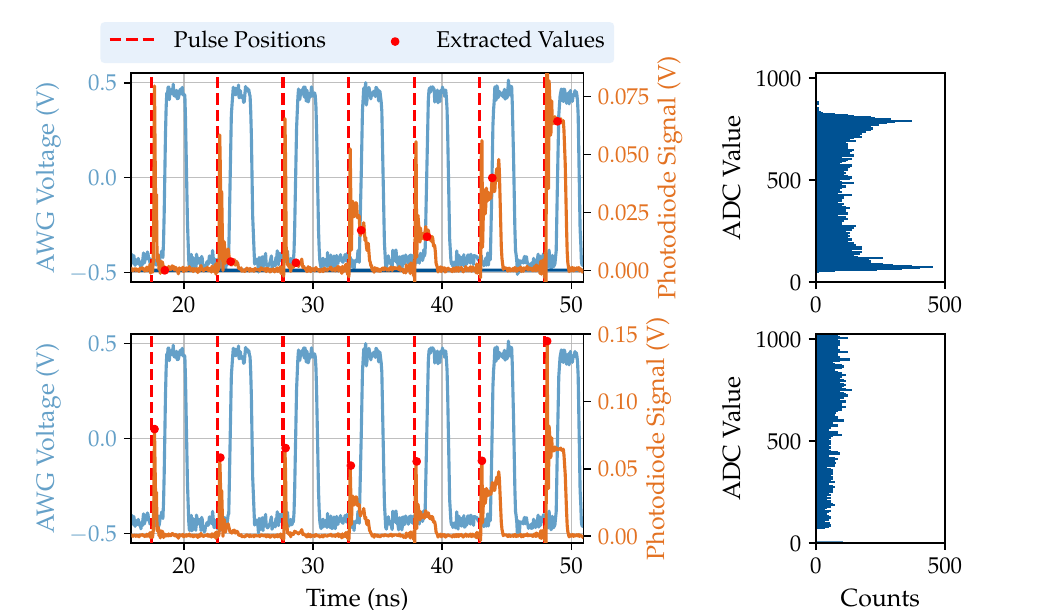}
    \caption{Pulse trace details from measurements with Laser 3. Top: The data points extracted after the relaxation oscillation regime give a arcsine-like distribution. Bottom: Taking the values in the relaxation oscillation period of the same pulses does not give a proper intensity distribution. The effect of the relaxation oscillation needs to be mitigated at high modulation frequencies.}
    \label{fig:relax_osci_bad_dist}
\end{figure}

At too short duty cycles, the criteria are not fulfilled as well. The reason for that are the highly fluctuating relaxation oscillations present at the beginning of the square pulses, causing multi-mode operation and other non-desired effects for interference measurements~\cite{Numai.2015}. Approaching \SI{}{\giga\hertz} modulation speeds, these oscillations decaying in the order of $\SI{}{\nano\second}$~\cite{Paraiso.2021} play an important role. If the pulses become too short, one can only measure intensities at points where these oscillations are still dominating. The comparison in Figure~\ref{fig:relax_osci_bad_dist} shows that the obtained intensity histograms depend heavily on the point of extraction within the pulse. Although a shorter duty cycle should lead to even better phase randomization as the off-time is longer, the data obtained does not guarantee underlying uniform phase distribution statistically, as no arcsine distribution is obtained. This does not mean that there is no phase randomization, but in order to extract quantum randomness from these statistics, more advanced analysis is necessary. In order to enable a higher random number generation speed with the proposed method, however, these relaxation oscillations should be suppressed. This works e.g. by injecting light into the cavity~\cite{Lang.1976} which might have negative effects on the phase randomization dynamics. Another approach to mitigate the influence of these relaxation oscillations connected with chirp of the laser light is to use spectral filters as suggested in~\cite{Shakhovoy.2021}.

Another temporal behavior was observed through the analysis of the two criteria. Comparing the three lasers in terms of acceptance at various duty cycle values, Laser 1 was found to be unable to produce any accepted data at a duty cycle of $\text{DC}=29/30$ which corresponds to an off-time of $\SI{0.17}{\nano\second}$. 
At the second largest value of $\text{DC}=26/30$ corresponding to an off-time of $\SI{0.67}{\nano\second}$ some successful measurements were obtained. So, it stops generating random numbers somewhere between these values. Pulsing with an an off-time of  $\SI{0.5}{\nano\second}$, a value between those two settings, with duty cycle $0.5$, one would end up with a pulse frequency of around $\SI{1}{\giga\hertz}$ where the laser begins to fail.
The reason for that was found by comparing the measurements to simulations of the experiment. Those simulations were performed with the rate equation model developed in~\cite{Bowers.1987} and the laser parameter extraction for it followed the procedure described in~\cite{Lovic.2021, Cartledge.1997}. The numerical method described in~\cite{Paraiso.2021} was used to simulate the rate equations. The experimental result could be reproduced in the simulations by applying a lowpass filter to the input current signal, suggesting that the electronics of the laser bias-T are responsible for the speed limit. Measurements of the amplitude modulation response also confirmed that by showing a sharply decreasing modulation response at a frequency of $\SI{1}{\giga\hertz}$. 
The datasheet of the bias-T obtained from the manufacturer indeed stated that there is a modulation limit at $\SI{1}{\giga\hertz}$. 
This shows that the developed criteria reflect the capabilities of the QRNG device in terms of its modulation behavior. 

As mentioned above, the fluctuations in autocorrelation are substantial and in the order of several $\SI{}{\decibel}$. For index $1$ in Figure~\ref{fig:laser3_autocor}, calculated from $100$ measurements at the same parameter setting, the mean of the gain-switching data is $\overline{\Gamma_1}=\SI{-20.8}{\decibel}$ but the standard deviation ranges from $\Gamma_{1,min} = \SI{-26.8}{\decibel}$ to $\Gamma_{1,max} = \SI{-18.4}{\decibel}$, therefore some measurements overshooting the boundary have to be expected. 
The mean and standard deviation of the statistical distance calculated from the same measurements at $T=\SI{25}{\celsius}$, $I_m=\SI{33.78}{\milli\ampere}$, $\text{MD}=0.51$, and $\text{DC}=0.57$ are $\overline{d_{stat}}=0.129$ and $\sigma_{d_{stat}}=0.004$. 
This deviation is in perfect agreement with the fluctuation simulation result presented in Figure~\ref{fig:stat_dist_vs_sample_size}. It can become relevant at operating conditions close to the statistical distance boundary, therefore parameters far enough from the boundary should be used  for stable operation of the QRNG. The parameter combination used to quantify this deviation, where the fluctuations of the statistical distance do not cause the dismiss of any extracted numbers, constitute an exemplary set of such parameters.

\begin{figure}
    \centering
    \includegraphics[width=1\linewidth]{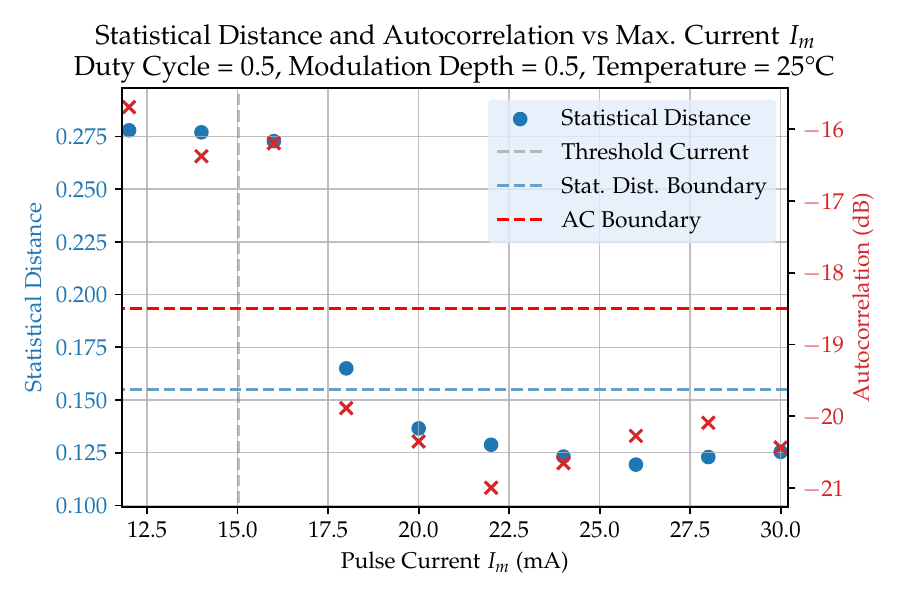}
    \caption{The statistical distance starts to decrease above the threshold current as the SNR improves. This happens gradually as the arcsine distribution is getting more and more prominent with the overlaying noise staying constant. The autocorrelation drops sharply to values below the boundary above a certain driving current as randomization does not happen gradually when varying the driving current.}
    \label{fig:stat_dist_vs_current}
\end{figure}

Studying the behavior of the criteria with respect to single varying parameters, the following additional trends are observed. First, the statistical distance decreases above the laser threshold and continues to decrease with increasing pulse current amplitude under gain-switching conditions. This is shown in Figure~\ref{fig:stat_dist_vs_current}. Improving the SNR is therefore considered to be the best way to improve the statistical quality of the QRNG. The autocorrelation behavior also shows a drop as soon as the signal surpasses the noise but does not decrease much further with increased noise. 
For the modulation depth there is no further improvement above a certain value, so a certain minimum must be guaranteed. Also for the duty cycle there is a certain threshold to overcome to reach pulse lengths exceeding the relaxation oscillation regime to reach a low statistical distance plateau.The pulse frequency limits for the lasers in terms of the minimum off-time required for phase-randomization could not be explored with the used setup. One would expect a gradual change in both criteria as soon as the phase diffusion width decreases below a certain value. 

\section{Conclusion}
A comprehensive evaluation framework for QRNG devices utilizing semiconductor laser phase noise was introduced. The established criteria effectively differentiate between secure and insecure output conditions based on the analyis of phase randomization and autocorrelation. The plausibility of the used criteria could be confirmed by the presented exemplary measurements. The developed tests are applicable for assessing QRNG quality in custom-built devices as required. It is crucial to emphasize that while the statistical randomness evaluated is vital for a dependable random number generator, true security necessitates further examination to prevent potential compromises by malicious entities. This responsibility lies with the operator. Furthermore, this testing suite does not assess the amount of extractable quantum randomness from the output, which requires comprehensive analyses of all classical noise sources. This task is challenging or potentially unfeasible due to the incomplete knowledge of all contributing sources, with the recent identification of phase-locking highlighting a new area of investigation~\cite{Shakhovoy.2025}. Nevertheless, our framework introduces a solid qualification routine which can be used by QRNG developers when selecting a suitable laser for their platform.

\section*{Funding}
The work reported in this manuscript was performed within the QuNET project, funded by the German Federal Ministry of Education and Research under the funding code 16KIS1265. The authors are responsible for the content of this publication.

\section*{Data availability} Data underlying the results presented in this paper are not publicly available at this time but may be obtained from the authors upon reasonable request.

\section*{Disclosures}
The authors declare no conflicts of interest.

\appendix
\section{Comparative Analysis of Laser Performance in Quantum Random Number Generation (QRNG)}
\label{app:comparison}
\subsection{Observations}
The qualification measurements of the three laser models gave insights into their suitability for QRNG applications. It was found that all three lasers can meet the required statistical distance and autocorrelation criteria under specific operating conditions, but their performance varies significantly. This section provides a short comparison of their behaviors, emphasizing their strengths and weaknesses in generating high-quality random numbers.

The absolute minimum statistical distance values achieved by the three lasers are as follows: $d_{stat,min} = \SI{0.103}{}$ for Laser 1, $\SI{0.125}{}$ for Laser 2, and $\SI{0.101}{}$ for Laser 3. These results indicate that the outputs of lasers 1 and 3 are closer to the ideal arcsine distribution under optimal conditions. The higher statistical distance for Laser 2 laser is caused by its comparatively low optical output power, leading to a lower SNR and distorting the intensity distribution.

\begin{figure}
    \centering
    \includegraphics[width=1\linewidth]{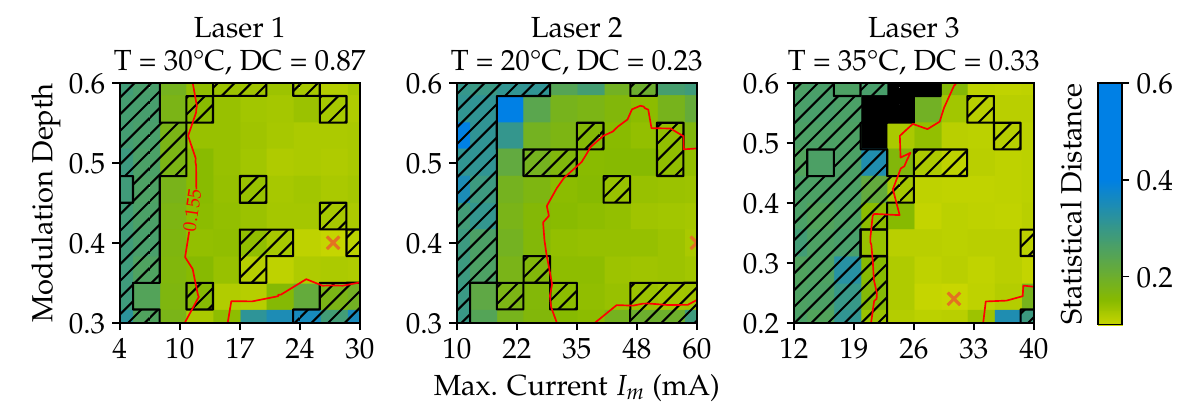}
    \caption{Three plots showing a certain plane in the parameter space of each laser. Constant parameters are given below the manufacturer title. For each laser, a parameter space plane containing the minimum statistical distance value achieved in the measurements is shown. The minimum values are indicated by the orange cross markers. The black squares represent measurement parameters where the fitting of the arcsine did not work, so that a statistical distance of $d_{\text{stat}}=1$ was obtained.}
    \label{fig:stat_dist_minima}
\end{figure}

\begin{figure}
    \centering
    \includegraphics[width=1\linewidth]{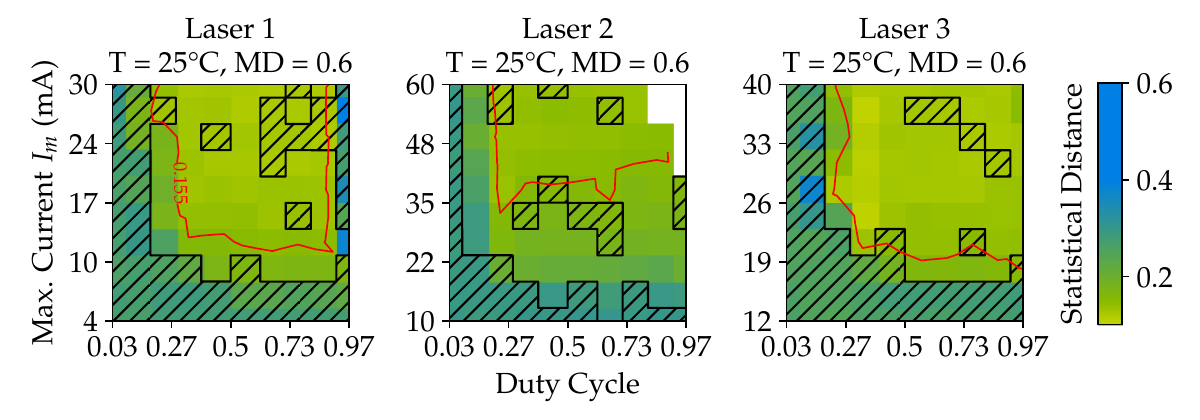}
    \caption{The statistical distance values plotted on the parameter planes spanned by the pulse peak current and the duty cycle parameters. At very low duty cycles, where the pulses are very short, the lasers do not produce accepted ouput. If only the first part of the pulse is used for interference, where the phase is unstable due to the strong relaxation oscillations, the retrieved distribution is not an arcsine. In that region, typical values of $d_{\text{stat}}\approx0.256$ are achieved with Laser 3, for example. Laser 1 shows insufficient randomization when the duty cycle approaches the maximum achievable value of $\text{DC}=29/30$. An application of Laser 1 for pulse frequencies above \SI{1}{\giga\hertz} is not recommended.}
    \label{fig:dc_vs_max_current_comparison}
\end{figure}

The statistical distance and autocorrelation criteria exclude similar parameter regions for lasers 1 and 3, as shown in Figures~\ref{fig:stat_dist_minima} and~\ref{fig:dc_vs_max_current_comparison}. For Laser 2, the overlap is weaker, likely due to its susceptibility to noise. This lase would need a loosened statistical distance boundary condition or reduced detection noise to meet both criteria in a wider window within its possible range of operating conditions, highlighting the need for tailored system design.

All lasers fail to meet the criteria at $\text{DC} = 1/30$ (pulses shorter than \SI{0.67}{\nano\second}) due to relaxation oscillations destabilizing phase and intensity. This causes Gaussian-like distributions instead of the desired arcsine, as illustrated in Figure~\ref{fig:relax_osci_bad_dist}. At $\text{DC} = 29/30$ (\SI{0.17}{\nano\second} off-time), Laser 1 cannot operate due to modulation limitations, while both both other models succeed at these DCs.

\begin{figure}
    \centering
    \includegraphics[width=1\linewidth]{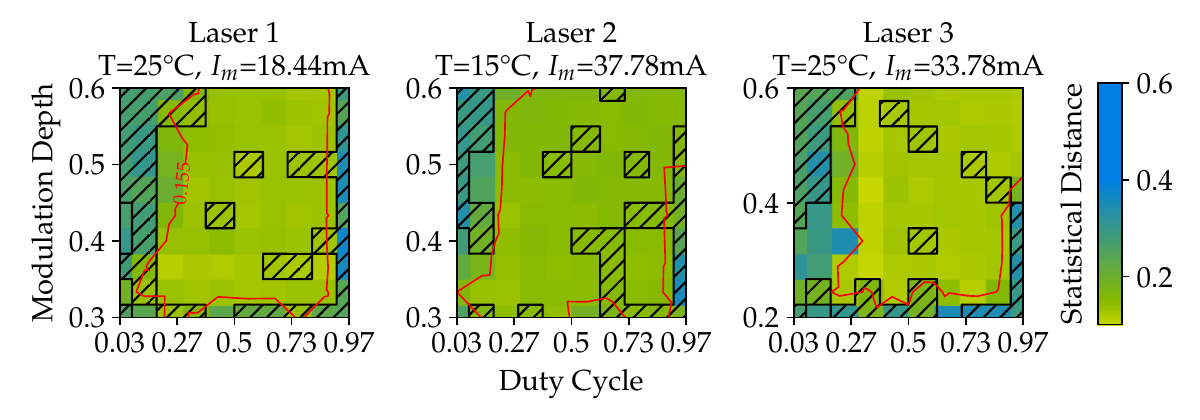}
    \caption{Statistical distance of measurements taken at constant temperatures and maximum currents. At very low duty cycles the statistical distance criterion is not fulfilled and at very high duty cycles a certain minimum modulation depth is necessary to achieve good operating conditions. This behavior shows that at high modulation speeds, it is important to lower the off-time bias current or even reverse bias the laser in this phase. This decreases the cavity field faster, leading to faster phase randomization.}
    \label{fig:stat_dist_md_vs_dc}
\end{figure}

At high DCs, models 2 and 3 need a minimum MD to lower the off-time driving current or even reverse bias the laser during that time, accelerating cavity field decay and phase diffusion, seen in Figure~\ref{fig:stat_dist_md_vs_dc}. As stated above, the modulation bandwidth of laser module 1 is limited so that it does not meet the requirements at extreme DCs.

Increasing the chip temperature raises the threshold currents of the lasers, reducing the required modulation depth for gain-switching. This trend is observed in all three lasers, validating the statistical distance boundary as a proxy for laser dynamics.


\subsection{System Limitations and Practical Considerations}
Laser 1's default Bias-T limits its modulation to \SI{1}{\giga\hertz}, making it unsuitable for high-speed QRNG application. As mentioned previously, rate equation simulations hinted and the manufacturer confirmed that this is an electronics issue, not the laser chip itself. Therefore, the limitation only holds for the specific electronics setup used. The other two models do not show this speed limitation.
The performance of Laser 2 is highly sensitive to detection noise, which can obscure the arcsine distribution. This underscores the need for a noise-optimized detection system tailored to the laser. Lasers 1 and 3 on the other hand demonstrate robustness to the noise of the system.

Laser 3 offers the best statistical distance and high duty cycle tolerance, making it ideal for high-speed QRNG applications even with non-optimized detection systems. Laser 1 is suitable for moderate-speed applications but is limited by its modulation electronics at high frequencies. Laser 2 requires careful system design to mitigate the influence of detection noise but can achieve QRNG operation at high duty cycles if these constraints are addressed.
For practical QRNG implementation, the choice of laser should balance statistical performance, modulation capabilities, and the influence of the system noise. Continuous monitoring of the statistical distance and the autocorrelation is essential to ensure compliance with defined security requirements for all lasers. 

\bibliography{paper} 

\end{document}